\documentclass[journal,12pt,onecolumn,draftclsnofoot,]{IEEEtran}
\usepackage[ruled, vlined]{algorithm2e}
\usepackage{algpseudocode}
\usepackage{amsmath}
\usepackage{amssymb}
\usepackage{array}
\usepackage{arydshln}
\usepackage{blkarray, bigstrut}
\usepackage{bm}
\usepackage{booktabs}
\usepackage{braket}
\usepackage{cite}
\usepackage{cuted}
\usepackage{enumitem}
\usepackage{graphicx}
\usepackage[mathscr]{euscript}
\usepackage{mathtools}
\usepackage{multirow}
\usepackage{multicol}
\usepackage{framed} 
\usepackage[framed]{ntheorem}
\usepackage{nicefrac}
\usepackage{pgfplots}
\usepackage{qtree}
\usepackage{adjustbox}
\usepackage{rotating}
\pgfplotsset{compat = 1.13}
\usepackage{stfloats}
\usepackage{subfig}
\usepackage{url}
\usepackage{tikz}
\usepackage{tkz-berge}
\usepackage[framemethod=TikZ]{mdframed}
\usetikzlibrary{patterns}
\usetikzlibrary{spy}
\usetikzlibrary{shapes, backgrounds,calc, snakes}
\usetikzlibrary{decorations.pathreplacing}
\usetikzlibrary{matrix}
\usepackage{fancybox}

\tikzstyle{vertex} = [circle, draw, inner sep = 0pt, minimum size = 10pt]

\newframedtheorem{frm-prop}{Property}

\definecolor{bblue}{rgb}{0.12392, 0.0490, 0.9588}
\definecolor{sskyblue}{rgb}{0.1529, 0.5882, 0.9216}
\definecolor{ggreen}{rgb}{0.5020, 0.7961, 0.3451}
\definecolor{yyellow}{rgb}{0.9765, 0.9804, 0.0784}

\definecolor{color0}{HTML}{FF0147}
\definecolor{color1}{HTML}{F400DC}
\definecolor{color2}{HTML}{BA0DFF}
\definecolor{color3}{HTML}{5700E8}
\definecolor{color4}{HTML}{0B03FF}
\definecolor{color5}{HTML}{0957F4}
\definecolor{color6}{HTML}{03B3FF}
\definecolor{color7}{HTML}{08E8DA}
\definecolor{color8}{HTML}{07FF8E}
\definecolor{color9}{HTML}{51FF0A}

\definecolor{p1}{rgb}{1, 0.0667, 0}
\definecolor{p2}{rgb}{1, 0.24, 0}
\definecolor{p3}{rgb}{1, 0.349, 0}
\definecolor{p4}{rgb}{1, 0.490, 0}
\definecolor{p5}{rgb}{1, 0.631, 0}
\definecolor{p6}{rgb}{1, 0.792, 0}
\definecolor{p7}{rgb}{1, 0.933, 0}
\definecolor{p8}{rgb}{1, 1, 0}
\definecolor{p9}{rgb}{1, 1, 0.5}
\definecolor{p10}{rgb}{1, 1, 0.8}

\begin{document}

\title{Enhanced C-V2X Mode-4 Subchannel Selection}

\author
{
	\IEEEauthorblockN{Luis F.~Abanto-Leon\IEEEauthorrefmark{2}, Arie Koppelaar\IEEEauthorrefmark{3}, Sonia Heemstra de Groot\IEEEauthorrefmark{4}} \\
	\IEEEauthorblockA{\IEEEauthorrefmark{2} \IEEEauthorrefmark{4} Eindhoven University of Technology}
	\IEEEauthorblockA{\IEEEauthorrefmark{3}NXP Semiconductors} \\ 
	\IEEEauthorrefmark{2}l.f.abanto@tue.nl,
	\IEEEauthorrefmark{3}arie.koppelaar@nxp.com,
	\IEEEauthorrefmark{4}sheemstradegroot@tue.nl,
}

\maketitle

\begin{abstract}
	In Release 14, the 3rd Generation Partnership Project (3GPP) introduced Cellular Vehicle--to--Everything (C-V2X) \textit{mode-4} as a novel disruptive technology to support sidelink vehicular communications in out--of--coverage scenarios. C-V2X \textit{mode-4} has been engineered to operate in a distributed manner, wherein vehicles autonomously monitor the received power across sidelink subchannels before selecting one for utilization. By means of such an strategy, vehicles attempt to $(i)$ discover and $(ii)$ reserve subchannels with low interference that may have the potential to maximize the reception likelihood of their own broadcasted safety messages. However, due to dynamicity of the vehicular environment, the subchannels optimality may fluctuate rapidly over time. As a consequence, vehicles are required to make a new selection every few hundreds of milliseconds. In consonance with 3GPP, the subchannel selection phase relies on the linear average of the perceived power intensities on each of the subchannels during a monitoring window. However, in this paper we propose a nonlinear power averaging phase, where the most up--to--date measurements are assigned higher priority via exponential weighting. We show through simulations that the overall system performance can be leveraged in both urban and freeway scenarios. Furthermore, the linear averaging can be considered as a special case of the exponentially-weighted moving average, ensuring backward compatibility with the standardized method. Finally, the 3GPP \textit{mode-4} scheduling approach is described in detail.
\end{abstract}


\IEEEpeerreviewmaketitle

\section{Introduction}
Cellular Vehicle--to--Everything (C-V2X) communications is one of the novel paradigms introduced by the 3rd Generation Partnership Project (3GPP) \cite{b1} in Release 14. C-V2X communications has been devised to be a dependable technology with the capability of displaying robustness in highly dynamic vehicular scenarios with varying densities, while satisfying stringent latency and reliability requirements. Thus, C-V2X has the potential to become a propitious asset that can be advantageously exploited in several application areas. For instance, vehicles with communication attributes may assist in preventing accidents and reducing the number of casualties \cite{b2}. Similarly, these communications capabilities can also be harnessed to optimize road traffic flow, which is anticipated to produce a plethora of positive impacts across several dimensions. 

Within C-V2X, two operation modalities are described: \textit{mode-3} and \textit{mode-4} \cite{b1}. The former one is a centralized scheme that relies on the availability of cellular infrastructure such as eNodeBs in order to distribute the available sidelink subchannels among the vehicles in coverage. eNodeBs may pursue multiple different criteria to accomplish such an objective. For instance, \cite{b3} considers a single sub-band setting where a sequential heuristic approach is proposed in order to maximize the reuse distance among vehicles broadcasting in the same subchannel, and thus leading to co-channel interference (CCI) reduction. On the other hand, \cite{b4} describes a multiple sub-band setting where maximization of the system sum-capacity is sought based on the subchannels signal--to--interference--plus--noise ratio (SINR) that vehicles report to eNodeBs. Furthermore, \cite{b5} extends the previous work including additional constraints where differentiated QoS requirements per vehicle are considered. Regardless of the optimization criteria, once an eNodeB has computed a suitable distribution of subchannels, vehicles will be notified of the resultant allocation via downlink. Thereupon, vehicles will engage in sidelink direct communications with their counterparts using the allotted resources. Contrastingly, \textit{mode-4} has been devised to operate in the absence of network coverage. In particular, such kind of situations might arise when cellular infrastructure has not been deployed in the area or when network coverage is not reliable enough to reckon on. As a consequence, vehicles will have to monitor the received power intensity on each subchannel and select a suitable one for utilization. Expressly, a vehicle will self-allocate a subchannel which may be unoccupied or experiences low interference in order to improve the likelihood of its own transmitted messages being received reliably. Although difficult to guarantee as a result of $(i)$ the distributedness of the scheme and $(ii)$ the unpredictability of channel fluctuations in the environment---by means of such an strategy of \textit{sensing before selecting}---not only do vehicles attempt to improve the reception probability of their messages but also strive not to impinge on the conditions of other subchannels being reserved by neighboring vehicles. In this manner, every vehicle in the system continuously endeavors to maintain an equilibrium point where interference can be minimized.

The comparative advantage of \textit{mode-3} is the more efficient utilization of subchannels that can be attained as a consequence of the humongous knowledge that eNodeBs can collect from all vehicles in coverage. Therefore, conflict-free subchannel assignments with minimal interference are realizable. Nonetheless, signaling between vehicles and eNodeBs via uplink/downlink constitutes a challenging task in terms of the rigorous latency exigencies that are required. Conversely, in \textit{mode-4} there is no need of a central controller to dictate assignments and therefore latency due to data collection is nonexistent. A noticeable drawback of \textit{mode-4} is the restricted local knowledge of each vehicle, which may cause the most satisfactory subchannels not to be always selected. Furthermore, due to incoordination, several vehicles may compete over the same subset of subchannels, and therefore leading to persistent conflicts and severe packet reception ratio (PRR) degradation.

In order to diminish the occurrences of conflicts, 3GPP standardized a semi-persistent scheduling (SPS) scheme whereby vehicles can reserve subchannels on a quasi-steady basis---in the order of a few hundreds of milliseconds---until re-scheduling is required. Thus, any receiving vehicle is capable of acquiring a degree of understanding on the subchannels utilization since short-term predictability is presumed. In dense scenarios, however, most of the subchannels might be under utilization and therefore vehicles must guide their selection based on the received power intensity, i.e. potential interference. When the reservation period of a subchannel has expired, a vehicle may have to process a new selection. This procedure consists in $(i)$ monitoring the received power on every subchannel during an observation window; $(ii)$ performing linear power averaging over such measurements in order to synthesize a metric representative of the interference level per subchannel; and finally $(iii)$ randomly selecting a subchannel among the best candidates. It has been proved by 3GPP through extensive simulations that such an strategy is consistent and robust enough to provide a fair basis of knowledge for vehicles to make a convenient selection while reducing the amount of concurrent conflicts. Given the necessity of further boosting reception reliability of messages, we propose a slight modification in the subchannel selection stage. Instead of relying on power linear averaging \cite{b1}, in the proposed approach the most recent measurements---within the observation window---are prioritized with higher weighting factors whereas received power intensities collected earlier in time are assigned lesser values. To wit, the most up-to-date values are more relevant for subchannel selection as these are representative of the current and short-term future utilization patterns. In addition, an optional feature allows vehicles to omit re-scheduling and reselect the currently reserved subchannel with some probability \cite{b6} \cite{b7}. We show through simulations that this profile attribute perfectly dovetails with the introduced exponentially-weighted moving average and can further boost the overall performance of the whole system.

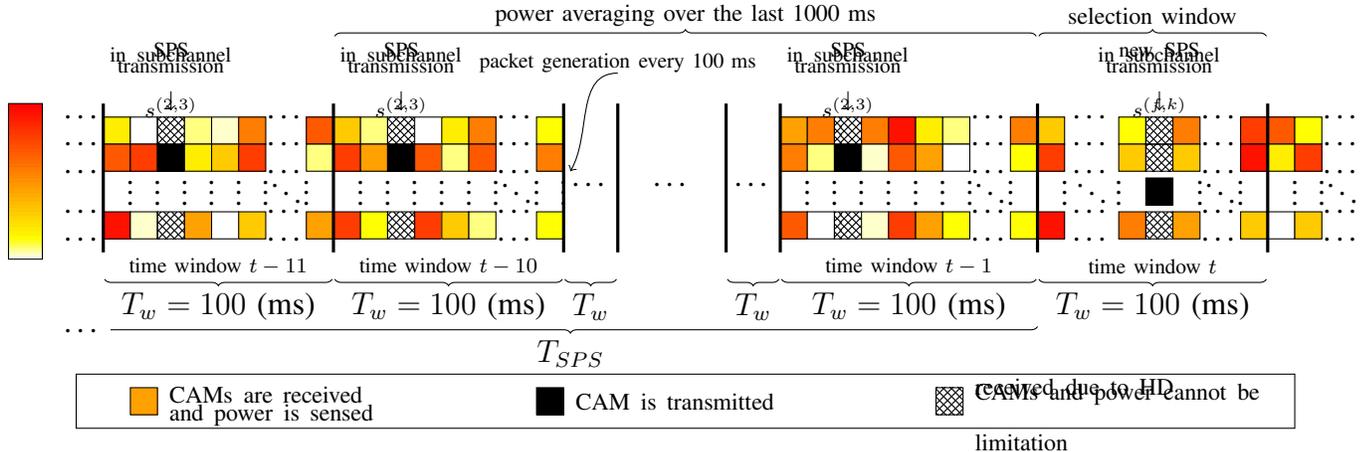
\begin{figure*}[!t]
	\centering
	\begin{tikzpicture}[scale = 0.9]
	
	\fill[top color=red, bottom color=yellow] (-1.0,0.2) rectangle (-0.5,-1.8) node[left]{};
	\fill[top color=yellow!, bottom color=white] (-1.0,-1.8) rectangle (-0.5,-2.1) node[left]{};
	\draw (-1.0,0.2) rectangle (-0.5,-2.1);
	
	\node at (0.1,0) {\dots};
	\node at (0.1,-0.4) {\dots};
	\node at (0.1,-0.8) {\dots};
	\node at (0.1,-1.4) {\dots};
	\node at (0.1,-1.8) {\dots};
	
	\node at (3.1,0) {\dots};
	\node at (3.1,-0.4) {\dots};
	\node at (3.1,-0.8) {\dots};
	\node at (3.1,-1.4) {\dots};
	\node at (3.1,-1.8) {\dots};
	
	\node at (6.5,0) {\dots};
	\node at (6.5,-0.4) {\dots};
	\node at (6.5,-0.8) {\dots};
	\node at (6.5,-1.4) {\dots};
	\node at (6.5,-1.8) {\dots};
	
	\node at (7.6,-1) {\dots};
	\node at (8.8,-1) {\dots};
	\node at (10,-1) {\dots};
	
	\node at (15,0) {\dots};
	\node at (15,-0.4) {\dots};
	\node at (15,-0.8) {\dots};
	\node at (15,-1.4) {\dots};
	\node at (15,-1.8) {\dots};
	
	\node at (13.5,0) {\dots};
	\node at (13.5,-0.4) {\dots};
	\node at (13.5,-0.8) {\dots};
	\node at (13.5,-1.4) {\dots};
	\node at (13.5,-1.8) {\dots};
	
	\node at (16.9,0) {\dots};
	\node at (16.9,-0.4) {\dots};
	\node at (16.9,-0.8) {\dots};
	\node at (16.9,-1.4) {\dots};
	\node at (16.9,-1.8) {\dots};
	
	\node at (18.7,0) {\dots};
	\node at (18.7,-0.4) {\dots};
	\node at (18.7,-0.8) {\dots};
	\node at (18.7,-1.4) {\dots};
	\node at (18.7,-1.8) {\dots};
	
	\node at (0.8,-1.0) {\vdots};
	\node at (1.2,-1.0) {\vdots};
	\node at (1.6,-1.0) {\vdots};
	\node at (2.0,-1.0) {\vdots};
	\node at (2.4,-1.0) {\vdots};
	\node at (2.8,-1.0) {\vdots};
	\node at (3.4,-1.0) {\vdots};
	\node at (4.2,-1.0) {\vdots};
	\node at (4.6,-1.0) {\vdots};
	\node at (5,-1.0) {\vdots};
	\node at (5.4,-1.0) {\vdots};
	\node at (5.8,-1.0) {\vdots};
	\node at (6.2,-1.0) {\vdots};
	\node at (6.8,-1.0) {\vdots};
	\node at (7.2,-1.0) {\vdots};
	\node at (10.8,-1.0) {\vdots};
	\node at (11.2,-1.0) {\vdots};
	\node at (11.6,-1.0) {\vdots};
	\node at (12.0,-1.0) {\vdots};
	\node at (12.4,-1.0) {\vdots};
	\node at (12.8,-1.0) {\vdots};
	\node at (13.2,-1.0) {\vdots};
	\node at (13.8,-1.0) {\vdots};
	\node at (14.6,-1.0) {\vdots};
	\node at (15.4,-1.0) {\vdots};
	\node at (16.6,-1.0) {\vdots};
	\node at (17.2,-1.0) {\vdots};
	\node at (18.0,-1.0) {\vdots};
	\node at (18.4,-1.0) {\vdots};
	
	\node at (3.1,-1.0) {$\ddots$};
	\node at (6.5,-1.0) {$\ddots$};
	\node at (13.5,-1.0) {$\ddots$};
	\node at (15.0,-1.0) {$\ddots$};
	\node at (16.9,-1.0) {$\ddots$};
	\node at (18.7,-1.0) {$\ddots$};
	
	\draw[fill=p7] (0.4,0) rectangle (0.8,-0.4);
	\draw[fill=p3] (0.4,-0.4) rectangle (0.8,-0.8);
	\draw[fill=p1] (0.4,-1.4) rectangle (0.8,-1.8);
	
	\draw[fill=white] (0.8,0) rectangle (1.2,-0.4);
	\draw[fill=p2] (0.8,-0.4) rectangle (1.2,-0.8);
	\draw[fill=p10] (0.8,-1.4) rectangle (1.2,-1.8);
	
	\draw[pattern=crosshatch, pattern color=black] (1.2,0) rectangle (1.6,-0.4);
	\draw[fill=black] (1.2,-0.4) rectangle (1.6,-0.8);
	\draw[pattern=crosshatch, pattern color=black] (1.2,-1.4) rectangle (1.6,-1.8);
	
	\draw[fill=p9] (1.6,0) rectangle (2,-0.4);
	\draw[fill=p7] (1.6,-0.4) rectangle (2,-0.8);
	\draw[fill=p5] (1.6,-1.4) rectangle (2,-1.8);
	
	\draw[fill=p10] (2,0) rectangle (2.4,-0.4);
	\draw[fill=p6] (2,-0.4) rectangle (2.4,-0.8);
	\draw[fill=white] (2,-1.4) rectangle (2.4,-1.8);
	
	\draw[fill=p4] (2.4,0) rectangle (2.8,-0.4);
	\draw[fill=p2] (2.4,-0.4) rectangle (2.8,-0.8);
	\draw[fill=p6] (2.4,-1.4) rectangle (2.8,-1.8);
	
	\draw[fill=p3] (3.4,0) rectangle (3.8,-0.4);
	\draw[fill=p9] (3.4,-0.4) rectangle (3.8,-0.8);
	\draw[fill=p5] (3.4,-1.4) rectangle (3.8,-1.8);
	
	\draw[fill=p6] (3.8,0) rectangle (4.2,-0.4);
	\draw[fill=p2] (3.8,-0.4) rectangle (4.2,-0.8);
	\draw[fill=p2] (3.8,-1.4) rectangle (4.2,-1.8);
	
	\draw[fill=p9] (4.2,0) rectangle (4.6,-0.4);
	\draw[fill=p5] (4.2,-0.4) rectangle (4.6,-0.8);
	\draw[fill=p8] (4.2,-1.4) rectangle (4.6,-1.8);
	
	\draw[pattern=crosshatch, pattern color=black] (4.6,0) rectangle (5,-0.4);
	\draw[fill=black] (4.6,-0.4) rectangle (5,-0.8);
	\draw[pattern=crosshatch, pattern color=black] (4.6,-1.4) rectangle (5,-1.8);
	
	\draw[fill=white] (5,0) rectangle (5.4,-0.4);
	\draw[fill=p3] (5,-0.4) rectangle (5.4,-0.8);
	\draw[fill=p2] (5,-1.4) rectangle (5.4,-1.8);
	
	\draw[fill=p7] (5.4,0) rectangle (5.8,-0.4);
	\draw[fill=p9] (5.4,-0.4) rectangle (5.8,-0.8);
	\draw[fill=p6] (5.4,-1.4) rectangle (5.8,-1.8);
	
	\draw[fill=p4] (5.8,0) rectangle (6.2,-0.4);
	\draw[fill=p3] (5.8,-0.4) rectangle (6.2,-0.8);
	\draw[fill=p9] (5.8,-1.4) rectangle (6.2,-1.8);
	
	\draw[fill=p8] (6.8,0) rectangle (7.2,-0.4);
	\draw[fill=p4] (6.8,-0.4) rectangle (7.2,-0.8);
	\draw[fill=p8] (6.8,-1.4) rectangle (7.2,-1.8);
	
	\draw[fill=p5] (10.4,0) rectangle (10.8,-0.4);
	\draw[fill=p4] (10.4,-0.4) rectangle (10.8,-0.8);
	\draw[fill=p3] (10.4,-1.4) rectangle (10.8,-1.8);
	
	\draw[fill=p4] (10.8,0) rectangle (11.2,-0.4);
	\draw[fill=p9] (10.8,-0.4) rectangle (11.2,-0.8);
	\draw[fill=white] (10.8,-1.4) rectangle (11.2,-1.8);
	
	\draw[pattern=crosshatch, pattern color=black] (11.2,0) rectangle (11.6,-0.4);
	\draw[fill=black] (11.2,-0.4) rectangle (11.6,-0.8);
	\draw[pattern=crosshatch, pattern color=black] (11.2,-1.4) rectangle (11.6,-1.8);
	
	\draw[fill=p4] (11.6,0) rectangle (12,-0.4);
	\draw[fill=p10] (11.6,-0.4) rectangle (12,-0.8);
	\draw[fill=p10] (11.6,-1.4) rectangle (12,-1.8);
	
	\draw[fill=p1] (12,0) rectangle (12.4,-0.4);
	\draw[fill=p3] (12,-0.4) rectangle (12.4,-0.8);
	\draw[fill=p2] (12,-1.4) rectangle (12.4,-1.8);
	
	\draw[fill=p7] (12.4,0) rectangle (12.8,-0.4);
	\draw[fill=p5] (12.4,-0.4) rectangle (12.8,-0.8);
	\draw[fill=p5] (12.4,-1.4) rectangle (12.8,-1.8);
	
	\draw[fill=p9] (12.8,0) rectangle (13.2,-0.4);
	\draw[fill=white] (12.8,-0.4) rectangle (13.2,-0.8);
	\draw[fill=p8] (12.8,-1.4) rectangle (13.2,-1.8);
	
	\draw[fill=p4] (13.8,0) rectangle (14.2,-0.4);
	\draw[fill=p8] (13.8,-0.4) rectangle (14.2,-0.8);
	\draw[fill=p8] (13.8,-1.4) rectangle (14.2,-1.8);
	
	\draw[fill=p6] (14.2,0) rectangle (14.6,-0.4);
	\draw[fill=p2] (14.2,-0.4) rectangle (14.6,-0.8);
	\draw[fill=p1] (14.2,-1.4) rectangle (14.6,-1.8);
	
	
	
	\draw[fill=p8] (15.4,0) rectangle (15.8,-0.4);
	\draw[fill=p6] (15.4,-0.4) rectangle (15.8,-0.8);
	\draw[fill=p4] (15.4,-1.4) rectangle (15.8,-1.8);
	
	\draw[fill=black] (15.8,-0.9) rectangle (16.2,-1.3);
	\draw[pattern=crosshatch, pattern color=black] (15.8,0) rectangle (16.2,-0.4);
	\draw[pattern=crosshatch, pattern color=black] (15.8,-0.4) rectangle (16.2,-0.8);
	\draw[pattern=crosshatch, pattern color=black] (15.8,-1.4) rectangle (16.2,-1.8);
	
	\draw[fill=p4] (16.2,0) rectangle (16.6,-0.4);
	\draw[fill=p6] (16.2,-0.4) rectangle (16.6,-0.8);
	\draw[fill=p5] (16.2,-1.4) rectangle (16.6,-1.8);
	
	\draw[fill=p2] (17.2,0) rectangle (17.6,-0.4);
	\draw[fill=p1] (17.2,-0.4) rectangle (17.6,-0.8);
	\draw[fill=p6] (17.2,-1.4) rectangle (17.6,-1.8);
	
	\draw[fill=p3] (17.6,0) rectangle (18.0,-0.4);
	\draw[fill=p7] (17.6,-0.4) rectangle (18.0,-0.8);
	\draw[fill=white] (17.6,-1.4) rectangle (18.0,-1.8);
	
	\draw[fill=p8] (18.0,0) rectangle (18.4,-0.4);
	\draw[fill=p2] (18.0,-0.4) rectangle (18.4,-0.8);
	\draw[fill=p6] (18.0,-1.4) rectangle (18.4,-1.8);
	
	\draw[very thick] (0.4,0.2) -- (0.4,-2);
	\draw[very thick] (3.8,0.2) -- (3.8,-2);
	\draw[very thick] (7.2,0.2) -- (7.2,-2);
	\draw[very thick] (8,0.2) -- (8,-2);
	\draw[very thick] (9.6,0.2) -- (9.6,-2);
	\draw[very thick] (10.4,0.2) -- (10.4,-2);
	\draw[very thick] (14.2,0.2) -- (14.2,-2);
	\draw[very thick] (17.6,0.2) -- (17.6,-2);
	
	\draw[decoration={brace, raise=5pt},decorate] (3.78,-2.2) -- node[right=6pt] {} (0.42,-2.2);
	\node at (2.1,-2.8) {$T_w = 100$ (ms)};
	\node at (2.1,-2.2) {\scriptsize time window $t-11$};
	
	\draw[decoration={brace, raise=5pt},decorate] (7.18,-2.2) -- node[right=6pt] {} (3.82,-2.2);
	\node at (5.5,-2.8) {$T_w = 100$ (ms)};
	\node at (5.5,-2.2) {\scriptsize time window $t-10$};
	
	\draw[decoration={brace, raise=5pt},decorate] (7.98,-2.2) -- node[right=6pt] {} (7.22,-2.2);
	\node at (7.6,-2.8) {$T_w$};
	
	\draw[decoration={brace, raise=5pt},decorate] (10.38,-2.2) -- node[right=6pt] {} (9.62,-2.2);
	\node at (10,-2.8) {$T_w$};
	
	\draw[decoration={brace, raise=5pt},decorate] (14.18,-2.2) -- node[right=6pt] {} (10.42,-2.2);
	\node at (12.3,-2.8) {$T_w = 100$ (ms)};
	\node at (12.3,-2.2) {\scriptsize time window $t-1$};
	
	\draw[decoration={brace, raise=5pt},decorate] (17.58,-2.2) -- node[right=6pt] {} (14.22,-2.2);
	\node at (15.9,-2.8) {$T_w = 100$ (ms)};
	\node at (15.9,-2.2) {\scriptsize time window $t$};
	
	\draw[decoration={brace, raise=5pt},decorate] (14.2,-2.9) -- node[right=6pt] {} (0.4,-2.9);
	\node at (7.3,-3.5) {$T_{SPS}$};
	\fill [white] (0.0,-2.7) rectangle (0.5,-3.3);
	\node at (0.1,-3.15) {\dots};
	
	\draw[decoration={brace, raise=5pt},decorate] (14.22,1) -- node[right=6pt] {} (17.58,1);
	\node at (15.9,1.5) {\footnotesize selection window};
	
	\draw[decoration={brace, raise=5pt},decorate] (3.82,1) -- node[right=6pt] {} (14.18,1);
	\node at (9,1.5) {\footnotesize power averaging over the last 1000 ms};
	
	\draw [->] (1.4,0.38) -- (1.4,0.1);
	\node at (1.4,1.0) [text width=2.3cm,align=center] {\scriptsize SPS};
	\node at (1.4,0.76) [text width=2.3cm,align=center] {\scriptsize transmission};
	\node at (1.4,0.52) [text width=2.3cm,align=center] {\scriptsize in subchannel $s^{(2,3)}$};
	
	\draw [->] (4.8,0.38) -- (4.8,0.1);
	\node at (4.8,1.0) [text width=2.3cm,align=center] {\scriptsize SPS};
	\node at (4.8,0.76) [text width=2.3cm,align=center] {\scriptsize transmission};
	\node at (4.8,0.52) [text width=2.3cm,align=center] {\scriptsize in subchannel $s^{(2,3)}$};
	
	\draw [->] (11.4,0.38) -- (11.4,0.1);
	\node at (11.4,1.0) [text width=2.3cm,align=center] {\scriptsize SPS};
	\node at (11.4,0.76) [text width=2.3cm,align=center] {\scriptsize transmission};
	\node at (11.4,0.52) [text width=2.3cm,align=center] {\scriptsize in subchannel $s^{(2,3)}$};
	
	\draw [->] (16.0,0.38) -- (16.0,0.1);
	\node at (16.0,1.0) [text width=2.3cm,align=center] {\scriptsize new SPS};
	\node at (16.0,0.76) [text width=2.3cm,align=center] {\scriptsize transmission};
	\node at (16.0,0.52) [text width=2.3cm,align=center] {\scriptsize in subchannel $s^{(f,k)}$};
	
	\draw[fill=p5] (0.8,-4.0) rectangle (1.2,-4.4);
	\node at (3.6,-4.1) [text width=4cm,align=left] {\footnotesize CAMs are received};
	\node at (3.6,-4.4) [text width=4cm,align=left] {\footnotesize and power is sensed};
	
	\draw[fill=black] (6.8,-4.0) rectangle (7.2,-4.4);
	\node at (9.6,-4.2) [text width=4cm,align=left] {\footnotesize CAM is transmitted};
	
	\draw[pattern=crosshatch, pattern color=black] (12.7,-4.0) rectangle (13.1,-4.4);
	\node at (15.5,-4.1) [text width=4cm,align=left] {\footnotesize CAMs and power cannot be};
	\node at (15.5,-4.4) [text width=4cm,align=left] {\footnotesize received due to HD limitation};
	
	\draw (0,-3.8) rectangle (18,-4.6);
	
	\node at (8,0.8) {\scriptsize packet generation every 100 ms};
	\draw[<-.] (7.3,-0.8) to [out=60, in=180] (8,0.65);
	
	
	\end{tikzpicture}	
	\caption{C-V2X \textit{mode-4} scheduling}
	\label{f1}
\end{figure*}

The paper is structured as follows. In Section II, the 3GPP SPS scheduling scheme for \textit{mode-4} communications is explained in detail. Section III describes the proposed exponential weighting variant for power averaging. Section IV is devoted to discussing simulation results obtained from real vehicular traces. Finally, Section V summarizes the conclusions of our work.

\section{3GPP Mode-4}
We have considered a 10 MHz intelligent transportation systems (ITS) channel for exclusive support of sidelink vehicular communications. Thus, the whole channel is divided into several time-frequency resource partitions---hereinafter called subchannels. Each has dimensions of one subframe (1 ms) in time and a number of resource blocks (RBs) in frequency. A subchannel is assumed to be capable of carrying a cooperative awareness message (CAM) and consists of two main components: data and control. The former one is also known as transport block (TB) and carries important information of each vehicle, e.g. position, speed, direction, etc. \cite{b8}. The latter portion is known as sidelink control information (SCI) \cite{b1} and transports relevant parameters---such as modulation and coding scheme (MCS), the number of resource blocks per subchannel, priority of the message, etc.---that will be employed for TB decoding and scheduling. In this work we have assumed a nominal message rate of $\Delta_{CAM} = 10$ Hz for all the vehicles in the system, and therefore the maximum amount of time divisions is 100. When a vehicle self-allocates a subchannel in a semi-persistent manner, it will periodically broadcast on such resource during $T_{SPS}$ ms, upon whose termination a new reservation will be required. For instance, Fig. \ref{f1} depicts the transmission and reception instances from a single vehicle perspective. It can be noticed that subchannel $s^{(2,3)}$---located at the intersection of sub-band $f=2$ and subframe $k=3$---is being persistently utilized every $T_w$ ms and such reservation pattern remains unchanged during $T_{SPS} / T_w$ consecutive time windows. Then---at the packet generation instance in the last reserved time window $t-1$---the vehicle selects its next SPS subchannel. In the following, we proceed to describe the 3GPP \textit{mode-4} scheduling scheme in more detail.

\subsection{Stage 1: Power Sensing}
Within a 10 MHz ITS channel, there exist $F$ sub-bands adjacent in frequency. Let $s^{(f,k)}$ denote the subchannel in sub-band $f$ (for $f=1, 2, \dots, F$) and subframe $k$ (for $k=1, 2, \dots, 100$) as depicted in Fig. \ref{f1}. Thus, $\mathcal{S} = \{ s^{(1,1)}, s^{(1,2)},\dots, s^{(F,100)} \}$ represents the complete set of $ | \mathcal{S} | =  100F$ subchannels for allocation. Since the value of $T_{SPS}=\{ 500, 600, \dots, 1500 \}$ is randomly drawn by each vehicle from a set of predetermined values \cite{b6}, the reservation period changes on a per vehicle basis, thus contributing to decorrelating the scheduling procedure among vehicles. During any specific time window $n$, a vehicle $v_i$ is persistently transmitting a CAM message of size $M_{CAM}$ bytes on a determined subchannel. Due to half-duplex PHY assumption, vehicle $v_i$ will be able to either transmit or receive at a time. Thus, as illustrated in Fig. \ref{f1}, the CAM messages in some subchannels and therefore their power intensities will not be received by the vehicle. 
\begin{figure*}[!t]
	\begin{equation} \label{e1}
	\small
	\varepsilon^{(n,f,k)}_i =
	\left\{
	\begin{array}{@{}ll@{}}
	\displaystyle \sum_{ \substack{{j=\{ u\mid v_u \in \mathcal{V}^{(n,k)}\}} \\  u \neq i } } I_{p} P_j \frac{G_t \cdot G_r}{\mathcal{X}_{ij}^{(n)} \cdot PL_{ij}^{(n)}} + P_{\sigma}, & \text{if}\ k = \{ m \mid {\mathcal{S}}_i^{(n)} \cap \{ s^{(1,m)},s^{(2,m)},\dots,s^{(F,m)} \} = \emptyset \} \\
	~~~~~~~~~~~~~~~~~~~~~~~~~~\infty, & \text{otherwise}
	\end{array}
	\right.
	\end{equation}
	\hrulefill
\end{figure*}
Let $\varepsilon^{(n,f,k)}_i$ denote the received power perceived by vehicle $v_i$ on a RB belonging to subchannel $s^{(f,k)}$ at any time window $n$. The power is computed as shown in (\ref{e1})\footnote{It is important to specify that the received power in this stage is calculated considering only the RBs pertaining to the reference signals within the TB. This metric is referred to as Physical Sidelink Shared Channel - Reference Signal Received Power (PSSCH-RSRP).}, where $\mathcal{V}^{(n,k)}$ represents the set of all the vehicles $v_j$ broadcasting on subchannels of subframe $k$ in time window $n$. On the other hand, ${\mathcal{S}}_i^{(n)}$ denotes the subset of subchannels that vehicle $v_i$ is utilizing in the current subframe $n$\footnote{The number of subchannels utilized by each vehicle can be 1 or 2 depending on whether retransmissions are enabled or not.}. Note from (\ref{e1}) that every subchannel belonging to a subframe---where vehicle $v_i$ has broadcasted---has had its power intensity set to $\infty$. The reason to this procedure is that the power could not be sensed due to half-duplex limitations and in order to preclude the selection of subchannels in unmonitored subframes, such power levels were assigned high values. The normalized RB transmit power of vehicle $v_j$ is represented by $P_j = P_T$, which is assumed to be the same for all units. The antenna gains of the transmitter and receiver are $G_t$ and $G_r$, respectively. The parameter $\mathcal{X}_{ij}^{(n)}$ is a log-normal random variable with standard deviation $\mathcal{X}_{\sigma}$ representing the shadowing experienced by the link between vehicles $v_i$ and $v_j$ at time window $n$. In addition, $PL_{ij}^{(n)} = \max \{PL_{ij}^{(n,{\text{free-space}})}, PL_{ij}^{(n,B_1)}\} $ depicts the path loss between $v_i$ and $v_j$. The first term represents the power attenuation based on the free-space model whereas the second term has been computed according to $\mathrm{WINNER+ UMi~(B\textsubscript{1})}$ \cite{b9} specifications. $P_{\sigma}$ represents the normalized noise floor per RB. $I_{p}$ is a factor that represents the influence of either co-channel interference (CCI) or in-band emissions (IBE) contributed by the any vehicles using subchannels of subframe $k$. $I_{p}$ is defined as the element in position ${\mid p-f+1 \mid}$ of a vector $\mathbf{I}$, where $p = \{ \tilde{f} \mid s^{(\tilde{f},k)} \in \mathcal{S}_j\}$. The elements of vector $\mathbf{I}$ represent the average energy level leaked from adjacent subchannels. For instance, in a configuration with $F=3$ sub-bands, $\mathbf{I} = [1~0.0047~0.0015]$ whereas for $F=4$, $\mathbf{I} = [1~0.0047~0.0015~0.0005]$. In-band emissions, path-loss and correlated shadowing have been modeled as specified in \cite{b10}. Thus, the average power $\tilde{\varepsilon}^{(n,f,k)}_i$ perceived by vehicle $v_i$ at time window $n$ is computed on the basis of measurements during the previous 10 time windows $\{ n-1, n-2, \dots, n-1000 \}$---i.e. a total of 1000 ms---where each subchannel will be averaged over 10 power samples.

\subsection{Stage 2: Subchannels Categorization}
Some subchannels will be excluded from selection based on the intensity of the (linear) average received power---obtained from the reference signals of the TB. Thus, if the average PSSCH-RSRP over the past 1000 ms exceeds a certain threshold $\gamma_{RSRP}$\footnote{This threshold is obtained considering the priorities of the CAM messages received in the subchannels and the priority of the message to be transmitted by $v_i$. In this work, we have assumed that the priority for all the messages is uniform and equal to 0. Thus, based on \cite{b7}, $\gamma_{RSRP} = -128$ dBm.}, those subchannels will be excluded as candidates for the new scheduling process. At this stage, the subchannels whose power could not be monitored, have been implicitly excluded as their power was set to $\infty$. If after this stage, the amount of allotable subchannels is less than 20\% of the initial number $| \mathcal{S} | = 100F$, the threshold $\gamma_{RSRP}$ is incremented by 3 dB and this process is iteratively repeated until the number of candidate subchannels is at least $ 0.2| \mathcal{S} |$. By means of increasing the threshold, the optimality of the candidate subchannels for scheduling is progressively relaxed. Thus, a vehicle becomes more permissive in including subchannels with slightly higher interference level at the expense of increasing the cardinality of the candidate set.

\subsection{Stage 3: Subchannel Selection}
Once the number of candidate subchannels is at least $ 0.2| \mathcal{S} |$, each vehicle $v_i$ will rank the subchannels in ascending order based on the linear average Received Signal Strength Indicator (RSSI)---which is computed across all the RBs of each subchannel. Thus, the selection process consists on each vehicle $v_i$ creating a set with the best $20F$ subchannels and then randomly choosing one for SPS transmission. In addition, an optional feature allows the vehicle---with probability $p_{keep}$---to skip re-scheduling and maintain the current subchannel \cite{b6} \cite{b7}. In such a case a new $T_{SPS}$ value will also be drawn.

\section{Exponentially-weighted Moving Average}
This procedure is applied to both \textit{Stages 2} and \textit{ Stage 3}, i.e. for PSSCH-RSRP and RSSI averaging. Although the linear average can provide a reliable impression of the interference degree, it has been noticed that by prioritizing the most recent measurements with higher weighting values, the performance of the system can be improved. Thus, the average received power of a RB---belonging to subchannel $s^{(f,k)}$---at time window $n$ is computed employing the exponentially-weighted measurements over the last 10 time windows, as shown in (\ref{e2})
\begin{equation} \label{e2}
	\small
	\tilde{\varepsilon}^{(n,f,k)}_i = {\displaystyle \Bigg[ \sum_{l=1}^{10} {\alpha}^{l} \Bigg]}^{-1}
	{\displaystyle \sum_{l=1}^{10} {\alpha}^{l} {\varepsilon}^{(n-l,f,k)}_i },
\end{equation}
where $\alpha \leq 1$ is an exponential weighting factor. When $\alpha = 1$, the procedures in \textit{Stage 2} and \textit{Stage 3} remain unaltered since the standardized linear average will be computed.

\section{Simulations}
In this section, we compare the standardized 3GPP scheduling method against the proposed variant. We evaluate two classes of vehicular scenarios---urban and freeway---and assess their performance in terms of the PRR using MATLAB. In addition, inspired by \cite{b11}, the types of error causing missed or undecodable packets are classified. For the urban case, a high vehicle density region of the \emph{TAPAS Cologne database} \cite{b12} was chosen, where an average number of 2000 vehicles over 60 seconds was observed. For the freeway case, a total number of 600 vehicles---distributed among 2 groups of 3 lanes per direction---with average density of 100 vehicles per kilometer was generated using Poisson point processes. In addition, the relevant parameters for the experiments are shown in Table \ref{t1}.
\begin{center}
	\begin{table}[!h]
		\centering
		\scriptsize
		\caption {Simulation parameters}
		\label{t1}
		\begin{tabular}{lccc}
			\toprule
			\multicolumn{1}{c}{\textbf{Description}} & \multicolumn{1}{c}{\textbf{Symbol}} & \multicolumn{1}{c}{\textbf{Value}} & \multicolumn{1}{c}{\textbf{Units}}\\
			\midrule
			Number of RBs per subchannel (per subframe) & - & 30 & - \\
			Number of sub-bands & $F$ & 3 & - \\
			Number of subchannels per sub-band & - & 100 & -\\
			Number of subchannels & - & 300 & -\\
			CAM message rate & $\Delta_{CAM}$ & 10 & Hz \\
			CAM size & $M_{CAM}$ & 190 & bytes\\
			MCS & - & 7 & - \\
			Transmit power per CAM & -& 23 & dBm \\
			Transmit power per RB & $P_T$ & 6.67 & mW \\
			Effective coded throughput (24 CRC bits) & $\rho$ & 0.9402 & bps$\slash$Hz \\
			Throughput loss coefficient \cite{b13} & $\lambda$ & 0.6 & - \\
			SINR threshold & $\gamma_{T}$ & 2.9293 & dB \\
			Distance between Tx and Rx & $D_x$ & 50-300 & m \\
			Scheduling period \cite{b7} & $T_{SPS}$ & 0.5-1.5 & s \\
			Antenna gain & $G_t, G_r$ & 3 & dB \\
			Shadowing standard deviation & $\mathcal{X}_{\sigma}$ & 7 & dB \\
			Shadowing correlation distance & - & 10 & m \\
			\bottomrule
		\end{tabular}
	\end{table}
	\vspace{-0.75cm}
\end{center}
\begin{figure}[!]
	\centering
	\begin{tikzpicture}[line cap=round,line join=round,x=2cm,y=2cm,spy using outlines={rectangle,lens={scale=3}, size=8cm, connect spies}]
	\begin{axis}[
	xmin = 50,
	xmax = 300,
	ymin = 0.79,
	width = 9.0cm,
	height = 5.5cm,
	xlabel={$D_x$ [meters]},
	x label style={align=center, font=\footnotesize,},
	ylabel = {Packet Reception (PRR\textsubscript{disk})},
	y label style={at={(-0.08,0.5)}, text width = 3.5cm, align=center, font=\footnotesize,},
	ytick = {0.5, 0.6, 0.7, 0.8, 0.9, 1.0},
	yticklabels = {0.5, 0.6, 0.7, 0.8, 0.9, 1.0},
	legend columns = 1,
	legend style={at={(0.008,0.01)},anchor=south west, font=\fontsize{6}{7}\selectfont, text width=2.4cm,text height=0.05cm,text depth=.ex, fill = none, align = left},
	]
	
	\addplot[color=black, mark = diamond*, mark options = {scale = 1.5, fill = green}, line width = 1pt] coordinates  
	{
		(50, 0.988178)
		(100, 0.977037)
		(150, 0.954630)
		(200, 0.918708)
		(250, 0.865511)
		(300, 0.798627)
	}; \addlegendentry{3GPP / $\alpha = 1$}

	\addplot[color = black, mark = pentagon*, mark options = {scale = 1.5, fill = color1, solid}, line width = 1pt] coordinates 
	{
		(50, 0.990219)
		(100, 0.982765)
		(150, 0.966774)
		(200, 0.936937)
		(250, 0.890013)
		(300, 0.826843)
	}; \addlegendentry{Proposed / $\alpha = 0.4$}

	\addplot[color=black, mark = square*, mark options = {fill = orange, solid}, line width = 1pt] coordinates
	{
		(50, 0.987951)
		(100, 0.979820)
		(150, 0.961992)
		(200, 0.931710)
		(250, 0.883990)
		(300, 0.820934)
	}; \addlegendentry{Proposed / $\alpha = 0.6$}

	\addplot[color=black, mark = triangle*, mark options = {scale = 1.5, fill = color6}, line width = 1pt] coordinates 
	{
		(50, 0.989673)
		(100, 0.980997)
		(150, 0.961476)
		(200, 0.928805)
		(250, 0.879522)
		(300, 0.814057)
	}; \addlegendentry{Proposed / $\alpha = 0.8$}

	\addplot[color=blue, mark options = {fill = yyellow}, line width = 1pt, style = solid] coordinates 
	{
		(50, 0.962143)
		(100, 0.939376)
		(150, 0.907496)
		(200, 0.865151)
		(250, 0.810263)
		(300, 0.745660)
	}; \addlegendentry{3GPP - Random / $\alpha = 1$}

	\addplot[color=black, mark options = {fill = yyellow}, line width = 1pt, style = densely dotted] coordinates 
	{
		(50, 0.938355)
		(100, 0.908862)
		(150, 0.873973)
		(200, 0.832216)
		(250, 0.779667)
		(300, 0.717359)
	}; \addlegendentry{3GPP - Greedy / $\alpha = 1$}
	\end{axis}
	
	\begin{axis}[
	xmin = 280,
	xmax = 300,
	ymin = 0.79,
	ymax = 0.84,
	width = 2.7cm,
	height = 3.2cm,
	ytick = {0.79, 0.80, 0.81, 0.82, 0.83, 0.84},
	yticklabels = {0.79, 0.80, 0.81, 0.82, 0.83, 0.84},
	xtick = {280,290,300},
	xticklabels = {280,290,300},
	tick label style={font=\fontsize{6}{6}\selectfont,},
	shift={(6.0cm,2.1cm)},axis background/.style={fill=white}
	]
	
	\addplot[color=black, mark = diamond*, mark options = {scale = 3, fill = green}, line width = 2pt] coordinates  
	{
		(50, 0.988178)
		(100, 0.977037)
		(150, 0.954630)
		(200, 0.918708)
		(250, 0.865511)
		(300, 0.798627)
	}; 
	
	\addplot[color = black, mark = pentagon*, mark options = {scale = 3, fill = color1, solid}, line width = 2pt] coordinates 
	{
		(50, 0.990219)
		(100, 0.982765)
		(150, 0.966774)
		(200, 0.936937)
		(250, 0.890013)
		(300, 0.826843)
	};

	\addplot[color=black, mark = square*, mark options = {scale = 2,fill = orange, solid}, line width = 2pt] coordinates
	{
		(50, 0.987951)
		(100, 0.979820)
		(150, 0.961992)
		(200, 0.931710)
		(250, 0.883990)
		(300, 0.820934)
	}; 
	
	\addplot[color=black, mark = triangle*, mark options = {scale = 3, fill = color6}, line width = 2pt] coordinates 
	{
		(50, 0.989673)
		(100, 0.980997)
		(150, 0.961476)
		(200, 0.928805)
		(250, 0.879522)
		(300, 0.814057)
	}; 
	
	\addplot[color=blue, mark options = {fill = yyellow}, line width = 1pt, style = solid] coordinates 
	{
		(50, 0.962143)
		(100, 0.939376)
		(150, 0.907496)
		(200, 0.865151)
		(250, 0.810263)
		(300, 0.745660)
	}; 
	
	\addplot[color=black, mark options = {fill = yyellow}, line width = 1pt, style = densely dotted] coordinates 
	{
		(50, 0.938355)
		(100, 0.908862)
		(150, 0.873973)
		(200, 0.832216)
		(250, 0.779667)
		(300, 0.717359)
	};

	\end{axis}
	
	\end{tikzpicture}
	\caption{PRR\textsubscript{disk} for an urban scenario with $p_{keep} = 0$}
	\label{f2}
	\vspace{-0.2cm}
\end{figure}
\begin{figure}[!]
	\centering
	\begin{tikzpicture}[line cap=round,line join=round,x=2cm,y=2cm,spy using outlines={rectangle,lens={scale=3}, size=8cm, connect spies}]
	\begin{axis}[
	xmin = 50,
	xmax = 300,
	ymin = 0.59,
	width = 9.0cm,
	height = 5.5cm,
	xlabel={$D_x$ [meters]},
	x label style={align=center, font=\footnotesize,},
	ylabel = {Packet Reception (PRR\textsubscript{ring})},
	y label style={at={(-0.08,0.5)}, text width = 3.5cm, align=center, font=\footnotesize,},
	ytick = {0.5, 0.6, 0.7, 0.8, 0.9, 1.0},
	yticklabels = {0.5, 0.6, 0.7, 0.8, 0.9, 1.0},
	legend columns = 1,
	legend style={at={(0.008,0.01)},anchor=south west, font=\fontsize{6}{7}\selectfont, text width=2.4cm,text height=0.05cm,text depth=.ex, fill = none, align = left},
	]
	
	\addplot[color=black, mark = diamond*, mark options = {scale = 1.5, fill = green}, line width = 1pt] coordinates 
	{
		(50, 0.988194)
		(100, 0.967375)
		(150, 0.919840)
		(200, 0.840963)
		(250, 0.724718)
		(300, 0.590403)
	}; \addlegendentry{3GPP / $\alpha = 1$}
	
	\addplot[color = black, mark = pentagon*, mark options = {scale = 1.5, fill = color1, solid}, line width = 1pt] coordinates 
	{
		(50, 0.990235)
		(100, 0.976301)
		(150, 0.941945)
		(200, 0.872362)
		(250, 0.765820)
		(300, 0.63018)
	}; \addlegendentry{Proposed / $\alpha = 0.4$}

	\addplot[color=black, mark = square*, mark options = {fill = orange, solid}, line width = 1pt] coordinates
	{
		(50, 0.987967)
		(100, 0.972768)
		(150, 0.934311)
		(200, 0.866172)
		(250, 0.757692)
		(300, 0.624631)
	}; \addlegendentry{Proposed / $\alpha = 0.6$}

	\addplot[color=black, mark = triangle*, mark options = {scale = 1.5, fill = color6}, line width = 1pt] coordinates 
	{
		(50, 0.989689)
		(100, 0.973473)
		(150, 0.931166)
		(200, 0.858096)
		(250, 0.749088)
		(300, 0.610251)
	}; \addlegendentry{Proposed / $\alpha = 0.8$}

	\addplot[color=blue, mark options = {fill = yyellow}, line width = 1pt, style = solid] coordinates 
	{
		(50, 0.962159)
		(100, 0.919629)
		(150, 0.857998)
		(200, 0.773505)
		(250, 0.664994)
		(300, 0.544540)
	}; \addlegendentry{3GPP - Random / $\alpha = 1$}

	\addplot[color=black, mark options = {fill = yyellow}, line width = 1pt, style = densely dotted] coordinates 
	{
		(50, 0.938370)
		(100, 0.883283)
		(150, 0.819802)
		(200, 0.741845)
		(250, 0.640587)
		(300, 0.523384)
	}; \addlegendentry{3GPP - Greedy / $\alpha = 1$}
	

	\end{axis}
	
	\begin{axis}[
	xmin = 280,
	xmax = 300,
	ymin = 0.58,
	ymax = 0.65,
	width = 2.7cm,
	height = 3.2cm,
	ytick = {0.58, 0.59, 0.60, 0.61, 0.62, 0.63, 0.64, 0.65},
	yticklabels = {0.58, 0.59, 0.60, 0.61, 0.62, 0.63, 0.64, 0.65},
	xtick = {280,290,300},
	xticklabels = {280,290,300},
	tick label style={font=\fontsize{6}{6}\selectfont,},
	shift={(6.0cm,2.1cm)},axis background/.style={fill=white}
	]
	
	\addplot[color=black, mark = diamond*, mark options = {scale = 3, fill = green}, line width = 2pt] coordinates 
	{
		(50, 0.988194)
		(100, 0.967375)
		(150, 0.919840)
		(200, 0.840963)
		(250, 0.724718)
		(300, 0.590403)
	};
	
	\addplot[color = black, mark = pentagon*, mark options = {scale = 3, fill = color1, solid}, line width = 2pt] coordinates 
	{
		(50, 0.990235)
		(100, 0.976301)
		(150, 0.941945)
		(200, 0.872362)
		(250, 0.765820)
		(300, 0.63018)
	}; 
	
	\addplot[color=black, mark = square*, mark options = {scale = 2, fill = orange, solid}, line width = 2pt] coordinates
	{
		(50, 0.987967)
		(100, 0.972768)
		(150, 0.934311)
		(200, 0.866172)
		(250, 0.757692)
		(300, 0.624631)
	}; 
	
	\addplot[color=black, mark = triangle*, mark options = {scale = 3, fill = color6}, line width = 2pt] coordinates 
	{
		(50, 0.989689)
		(100, 0.973473)
		(150, 0.931166)
		(200, 0.858096)
		(250, 0.749088)
		(300, 0.610251)
	}; 
	
	\addplot[color=blue, mark options = {fill = yyellow}, line width = 1pt, style = solid] coordinates 
	{
		(50, 0.962159)
		(100, 0.919629)
		(150, 0.857998)
		(200, 0.773505)
		(250, 0.664994)
		(300, 0.544540)
	}; 
	
	\addplot[color=black, mark options = {fill = yyellow}, line width = 1pt, style = densely dotted] coordinates 
	{
		(50, 0.938370)
		(100, 0.883283)
		(150, 0.819802)
		(200, 0.741845)
		(250, 0.640587)
		(300, 0.523384)
	}; 
	
	\end{axis}
	
	\end{tikzpicture}
	\caption{PRR\textsubscript{ring} for an urban scenario with $p_{keep} = 0$}
	\label{f3}
	\vspace{-0.2cm}
\end{figure}
\begin{figure}[!]
	\centering
	\begin{tikzpicture}[line cap=round,line join=round,x=2cm,y=2cm,spy using outlines={rectangle,lens={scale=3}, size=8cm, connect spies}]
	\begin{axis}[
	xmin = 50,
	xmax = 300,
	ymin = 0.59,
	width = 9.0cm,
	height = 5.5cm,
	xlabel={$D_x$ [meters]},
	x label style={align=center, font=\footnotesize,},
	ylabel = {Packet Reception (PRR\textsubscript{ring})},
	y label style={at={(-0.08,0.5)}, text width = 3.5cm, align=center, font=\footnotesize,},
	ytick = {0.5, 0.6, 0.7, 0.8, 0.9, 1.0},
	yticklabels = {0.5, 0.6, 0.7, 0.8, 0.9, 1.0},
	legend columns = 1,
	legend style={at={(0.008,0.01)},anchor=south west, font=\fontsize{6}{7}\selectfont, text width=2.7cm,text height=0.05cm,text depth=.ex, fill = none, align = left},
	]
	
	\addplot[color=black, mark = diamond*, mark options = {scale = 1.5, fill = green}, line width = 1pt] coordinates 
	{
		(50, 0.991757)
		(100, 0.975477)
		(150, 0.934869)
		(200, 0.861953)
		(250, 0.747246)
		(300, 0.611382)
	}; \addlegendentry{3GPP / $\alpha = 1$}

	\addplot[color = black, mark = pentagon*, mark options = {scale = 1.5, fill = color1, solid}, line width = 1pt] coordinates 
	{
		(50, 0.992885)
		(100, 0.980172)
		(150, 0.947666)
		(200, 0.882971)
		(250, 0.774143)
		(300, 0.641073)
	}; \addlegendentry{Proposed / $\alpha = 0.4$}
	
	\addplot[color=black, mark = square*, mark options = {fill = orange, solid}, line width = 1pt] coordinates
	{
		(50, 0.991254)
		(100, 0.979856)
		(150, 0.944664)
		(200, 0.874916)
		(250, 0.765537)
		(300, 0.633015)
	}; \addlegendentry{Proposed / $\alpha = 0.6$}
	
	\addplot[color=black, mark = triangle*, mark options = {scale = 1.5, fill = color6}, line width = 1pt] coordinates 
	{
		(50, 0.991047)
		(100, 0.976652)
		(150, 0.937514)
		(200, 0.866139)
		(250, 0.756782)
		(300, 0.623254)
	}; \addlegendentry{Proposed / $\alpha = 0.8$}
	
	\addplot[color=blue, mark options = {fill = yyellow}, line width = 1pt, style = solid] coordinates 
	{
		(50, 0.962343)
		(100, 0.921947)
		(150, 0.859632)
		(200, 0.777787)
		(250, 0.665269)
		(300, 0.543565)
	}; \addlegendentry{3GPP - Random / $\alpha = 1$}
	
	\addplot[color=black, mark options = {fill = yyellow}, line width = 1pt, style = densely dotted] coordinates 
	{
		(50, 0.961819)
		(100, 0.919018)
		(150, 0.874812)
		(200, 0.807803)
		(250, 0.700216)
		(300, 0.576681)
	}; \addlegendentry{3GPP - Greedy / $\alpha = 1$}

	\addplot[color=red, mark options = {fill = yyellow}, line width = 1pt, style = densely dotted] coordinates 
	{
		(50, 0.988194)
		(100, 0.967375)
		(150, 0.919840)
		(200, 0.840963)
		(250, 0.724718)
		(300, 0.590403)
	}; \addlegendentry{3GPP / $\alpha = 1$ ($p_{keep} = 0$)}	
	\end{axis}
	
	\begin{axis}[
	xmin = 280,
	xmax = 300,
	ymin = 0.6,
	ymax = 0.66,
	width = 2.7cm,
	height = 3.2cm,
	ytick = {0.60, 0.61, 0.62, 0.63, 0.64, 0.65, 0.66},
	yticklabels = {0.60, 0.61, 0.62, 0.63, 0.64, 0.65, 0.66},
	xtick = {280,290,300},
	xticklabels = {280,290,300},
	tick label style={font=\fontsize{6}{6}\selectfont,},
	shift={(6.0cm,2.1cm)},axis background/.style={fill=white}
	]
	
	\addplot[color=black, mark = diamond*, mark options = {scale = 3, fill = green}, line width = 2pt] coordinates 
	{
		(50, 0.991757)
		(100, 0.975477)
		(150, 0.934869)
		(200, 0.861953)
		(250, 0.747246)
		(300, 0.61138)
	}; 
	
	\addplot[color = black, mark = pentagon*, mark options = {scale = 3, fill = color1, solid}, line width = 2pt] coordinates 
	{
		(50, 0.992885)
		(100, 0.980172)
		(150, 0.947666)
		(200, 0.882971)
		(250, 0.774143)
		(300, 0.641073)
	}; 
	
	\addplot[color=black, mark = square*, mark options = {scale = 2,fill = orange, solid}, line width = 2pt] coordinates
	{
		(50, 0.991254)
		(100, 0.979856)
		(150, 0.944664)
		(200, 0.874916)
		(250, 0.765537)
		(300, 0.633015)
	}; 
	
	\addplot[color=black, mark = triangle*, mark options = {scale = 3, fill = color6}, line width = 2pt] coordinates 
	{
		(50, 0.991047)
		(100, 0.976652)
		(150, 0.937514)
		(200, 0.866139)
		(250, 0.756782)
		(300, 0.623254)
	}; 
	
	\addplot[color=blue, mark options = {fill = yyellow}, line width = 1pt, style = solid, line width = 2pt] coordinates 
	{
		(50, 0.962343)
		(100, 0.921947)
		(150, 0.859632)
		(200, 0.777787)
		(250, 0.665269)
		(300, 0.543565)
	}; 
	
	\addplot[color=black, mark options = {fill = yyellow}, line width = 1pt, style = densely dotted, line width = 2pt] coordinates 
	{
		(50, 0.961819)
		(100, 0.919018)
		(150, 0.874812)
		(200, 0.807803)
		(250, 0.700216)
		(300, 0.576681)
	}; 
	
	\addplot[color=red, mark options = {fill = yyellow}, line width = 1pt, style = densely dotted, line width = 2pt] coordinates 
	{
		(50, 0.988194)
		(100, 0.967375)
		(150, 0.919840)
		(200, 0.840963)
		(250, 0.724718)
		(300, 0.590403)
	}; 	
	
	\end{axis}

	\end{tikzpicture}
	\caption{PRR\textsubscript{ring} for an urban scenario with $p_{keep} = 0.2$}
	\label{f4}
	\vspace{-0.2cm}
\end{figure}

Fig. \ref{f2} and Fig. \ref{f3} compare the performance of the two approaches in a urban scenario with $p_{keep} = 0$. The type of PRR shown in Fig. \ref{f2}, i.e. PRR\textsubscript{disk}, represents the mainstream metric that counts all the successfully decoded packets considering every vehicle within a disk of radius $D_x$ from the transmitter. The second metric, namely PRR\textsubscript{ring}, considers only the vehicles located in the ring between $D_x$ and $D_x - 50$. This latter metric was introduced by 3GPP in \cite{b10} in order to evaluate the performance of a specific target group. In addition, the performance curves for greedy and random selection have been included. In the former case, several vehicles experiencing similar subchannel conditions may unknowingly select the same resources; thus leading to an increased amount of collisions. In the latter case, the quantity of packets colliding decreases since the whole set of candidate subchannels is $\mathcal{S}$. Nevertheless, subchannels with high interference may be selected and thus impinging on the PRR. The performance of both random and greedy approaches is suboptimal compared to the standardized method. On the other hand, we can observe that the proposed variant can produce PRR improvement in the near-field and far-field of each transmitting vehicle. This behavior is observed in both kinds of PRR, where the gains for $\alpha = 0.4$ at $D_x = 300$ are 2.82\% and 3.98\%, respectively. It was observed that values smaller than 0.4 tended to decrease performance, to extents lower than when $\alpha = 1$. Fig. \ref{f4} shows the PRR\textsubscript{ring} performance when $p_{keep} = 0.2$. It can be observed that this optional feature has the potential to boost the performance when compared to the case with $p_{keep} = 0$. However, when $p_{keep} > 0.2$ the dynamism due subchannel allocation changes is insufficient and therefore the PRR suffers degradation due to stasis. Furthermore, the proposed approach dovetails suitably with parameter $p_{keep}$ and their joint utilization is advantageous for enhancing the system performance. It can also be noticed that the random allocation is unaffected by $p_{keep}$ whereas the greedy selection is greatly benefited to the extent that it surpasses the former approach.

The packet errors have been classified in several categories as shown in Table \ref{t2} and Table \ref{t3}. In order of hierarchy, the following mutually exclusive classes are recognized: $(i)$ half-duplex impairment at subchannel level (HD-SC); $(ii)$ half-duplex impairment at subframe level (HD-SF); undecodable packets due to $(iii)$ propagation, $(iv)$ co-channel interference and $(v)$ in-band emissions. Basing our observations on PRR\textsubscript{ring}, in urban scenarios the most detrimental cause for lost packets is CCI while IBE and propagation are responsible for most of the remaining errors. Unreceived packets due to half-duplex (HD-SC and HD-SF) amount less than 1\%. On the other hand---in the freeway scenario---because the channel parameters have not been changed with respect to the urban case, we can observe a similar amount of lost packets due to propagation. However, in this case CCI is more relevant since the distribution of vehicles is more condensed; therefore the subchannel reuse distance among vehicles is shorter than in the urban scenario. As a consequence of vehicles being closely packed, the power leakage due to IBE is also more impactul and affects the PRR comparatively more severe than in the urban case.

\textbf{Note:} Across all the simulations, the PRR is computed checking whether every pair of vehicles $v_i$ and $v_j$ is within the awareness distance $D_x$ or not. If affirmative, the SINR $\gamma^{(f,k)}_{ij}$ experienced by $v_i$ upon reception of a packet transmitted by $v_j$ on subchannel $s^{(f,k)}$ is compared against a threshold $\gamma_{T} = 10 \cdot \log_{10}(2^{\rho/\lambda}-1)$ \cite{b13}, which is derived from the parameters in Table \ref{t1}. It is assumed that a message can be correctly decoded if its SINR is larger than $\gamma_{T}$.

\begin{table*}[!t]
	\centering
	\scriptsize
	\caption {Classification in percentage of missed/undecodable packets - Urban scenario with $\alpha = 1$ and $p_{keep} = 0$}
	\label{t2}
	\begin{tabular}{cccccccccccccccccccccccccc}
		\toprule
		\multirow{2}{*}{\textbf{Distance}} & 
		\multicolumn{1}{c}{\textbf{PRR}} & 
		\multicolumn{1}{c}{\textbf{HD-SF}} & 
		\multicolumn{1}{c}{\textbf{HD-SC}} & 
		\multicolumn{1}{c}{\textbf{Propagation}} &
		\multicolumn{1}{c}{\textbf{CCI}} &
		\multicolumn{1}{c}{\textbf{IBE}} &
		\multicolumn{1}{c}{\textbf{PRR}} & 
		\multicolumn{1}{c}{\textbf{HD-SF}} & 
		\multicolumn{1}{c}{\textbf{HD-SC}} & 
		\multicolumn{1}{c}{\textbf{Propagation}} &
		\multicolumn{1}{c}{\textbf{CCI}} &
		\multicolumn{1}{c}{\textbf{IBE}} & \\
									&
		\multicolumn{1}{c}{(Disk)} 	& 
		\multicolumn{1}{c}{(Disk)} 	& 
		\multicolumn{1}{c}{(Disk)} 	& 
		\multicolumn{1}{c}{(Disk)} 	&
		\multicolumn{1}{c}{(Disk)} 	&
		\multicolumn{1}{c}{(Disk)} 	&
		\multicolumn{1}{c}{(Ring)} 	& 
		\multicolumn{1}{c}{(Ring)} 	& 
		\multicolumn{1}{c}{(Ring)} 	& 
		\multicolumn{1}{c}{(Ring)} 	&
		\multicolumn{1}{c}{(Ring)} 	&
		\multicolumn{1}{c}{(Ring)} 	\\
				 	
		\midrule
		50 & 98.8194 & 0.1262 & 0.1050 & 0.0000 & 0.8664 & 0.0830 & 98.8194 & 0.1262 & 0.1050 & 0.0000 & 0.8664 & 0.0830 \\
		100 & 97.7037 & 0.2167 & 0.1093 & 0.0031 & 1.5919 & 0.3753 & 96.7375 & 0.2952 & 0.1131 & 0.0058 & 2.2195 & 0.6289 \\
		150 & 95.4630 & 0.3354 & 0.1076 & 0.0799 & 2.9353 & 1.0788 & 91.9840 & 0.5197 & 0.1036 & 0.1990 & 5.0226 & 2.1711 \\
		200 & 91.8708 & 0.4291 & 0.1025 & 0.6057 & 5.0871 & 1.9048 & 84.0963 & 0.6320 & 0.0916 & 1.7436 & 9.7441 & 3.6924 \\
		250 & 86.5511 & 0.5163 & 0.1017 & 2.3065 & 7.8852 & 2.6392 & 72.4718 & 0.7469 & 0.1005 & 6.8081 & 15.2899 & 4.5828 \\
		300 & 79.8627 & 0.5623 & 0.1148 & 5.5492 & 10.7124 & 3.1986 & 59.0403 & 0.7051 & 0.1553 & 15.6443 & 19.5148 & 4.9402 \\
		
		\bottomrule
	\end{tabular}
\end{table*}

\begin{table*}[!t]
	\centering
	\scriptsize
	\caption {Classification in percentage of missed/undecodable packets - Freeway scenario with $\alpha = 1$ and $p_{keep} = 0$}
	\label{t3}
	\begin{tabular}{cccccccccccccccccccccccccc}
		\toprule
		\multirow{2}{*}{\textbf{Distance}} & 
		\multicolumn{1}{c}{\textbf{PRR}} & 
		\multicolumn{1}{c}{\textbf{HD-SF}} & 
		\multicolumn{1}{c}{\textbf{HD-SC}} & 
		\multicolumn{1}{c}{\textbf{Propagation}} &
		\multicolumn{1}{c}{\textbf{CCI}} &
		\multicolumn{1}{c}{\textbf{IBE}} &
		\multicolumn{1}{c}{\textbf{PRR}} & 
		\multicolumn{1}{c}{\textbf{HD-SF}} & 
		\multicolumn{1}{c}{\textbf{HD-SC}} & 
		\multicolumn{1}{c}{\textbf{Propagation}} &
		\multicolumn{1}{c}{\textbf{CCI}} &
		\multicolumn{1}{c}{\textbf{IBE}} & \\
		&
		\multicolumn{1}{c}{(Disk)} 	& 
		\multicolumn{1}{c}{(Disk)} 	& 
		\multicolumn{1}{c}{(Disk)} 	& 
		\multicolumn{1}{c}{(Disk)} 	&
		\multicolumn{1}{c}{(Disk)} 	&
		\multicolumn{1}{c}{(Disk)} 	&
		\multicolumn{1}{c}{(Ring)} 	& 
		\multicolumn{1}{c}{(Ring)} 	& 
		\multicolumn{1}{c}{(Ring)} 	& 
		\multicolumn{1}{c}{(Ring)} 	&
		\multicolumn{1}{c}{(Ring)} 	&
		\multicolumn{1}{c}{(Ring)} 	\\
		
		\midrule
		50  & 97.8500 & 0.0911 & 0.2265 & 0.0000 & 1.5033 & 0.3291 & 97.8500 & 0.0911 & 0.2265 & 0.0000 & 1.5033 & 0.3291 \\ 
		100 & 94.8317 & 0.0940 & 0.4026 & 0.0032 & 3.2755 & 1.3930 & 93.1256 & 0.0957 & 0.5021 & 0.0050 & 4.2773 & 1.9943 \\
		150 & 91.2680 & 0.0999 & 0.4870 & 0.0611 & 5.2521 & 2.8318 & 84.4430 & 0.1113 & 0.6486 & 0.1721 & 9.0375 & 5.5875 \\
		200 & 87.1069 & 0.1088 & 0.5463 & 0.4599 & 7.4548 & 4.3233 & 73.9374 & 0.1369 & 0.7341 & 1.7222 & 14.4259 & 9.0434 \\
		250 & 82.6517 & 0.1193 & 0.5834 & 1.5996 & 9.4906 & 5.5553 & 62.7991 & 0.1661 & 0.7489 & 6.6780 & 18.5624 & 11.0456 \\
		300 & 78.0224 & 0.1350 & 0.6020 & 3.6732 & 11.1518 & 6.4157 & 51.2963 & 0.2256 & 0.7088 & 15.6445 & 20.7422 & 11.3826 \\
		
		\bottomrule
	\end{tabular}
\end{table*}

\section{Conclusion}
We have proposed an alternative to the standardized linear power averaging procedure---for PSSCH-RSRP and RSSI resource blocks---which has shown a positive impact in terms of PRR performance. In addition, we have shown through simulations that in two different environments, namely urban and freeway, the proposed variant is capable of excelling the standardized method. It should be noted that such a gain is only due to an improved management of subchannel selection in the scheduling procedure since no other features such as congestion control were introduced. For this reason, we foresee the potential of this modification to be combined with more advanced processes and functionalities. In addition, the proposed technique can be adaptive and tune its own parameters based on the sensed subchannels congestion.

\end{document}